Co-intercalation of Ammonia molecules between FeSe layers in Ba-doped FeSe superconductor

Seiji Shibasaki[1], Keishi Ashida[1], Yuki Takahei[1], Keitaro Tomita[1], Reiji Kumai[2] and Takashi Kambe[1,*]

[1]*Department of Physics, Faculty of Science, Okayama University, Okayama 700-8530, Japan*
[2]*Condensed Matter Research Center (CMRC) and Photon Factory, Institute of Materials Structure Science, High Energy Accelerator Research Organization (KEK), Tsukuba, 305-0801, Japan*

We have synthesized a Ba-doped FeSe superconductor using liquid ammonia, and investigated in detail the structure and composition of the Ba-FeSe phase. In the superconducting phase, both the $T_c$ and the shielding fraction are almost unchanged even if the nominal composition of Ba is varied in the region from 0.3 to 0.7. The optimized $T_c$ is 38.8 K. We have determined the structure of the superconducting phase with X-ray powder diffraction (using synchrotron radiation) and here describe the co-intercalation of Ba and $NH_3$ molecules between the FeSe layers.

PACS number: [74.70.Xa, 74.25.-q, 61.05.cp ]

* Corresponding author; kambe@science.okayama-u.ac.jp

Iron selenide (FeSe) shows a superconducting transition at 8 K [1]. The crystal structures of Fe-chalcogenides consist of single conduction layers, which is the simplest structure among iron-based superconductors. The Fermi surface structure of Fe-chalcogenides is similar to that of FeAs-based materials [2]. Thus, Fe-chalcogenides are widely regarded as a key material in the effort to understand the mechanism of superconductivity in iron-based materials. Moreover, the intercalation of alkali and alkali-earth metals in FeSe produces new high-$T$c phases, with $T$c values up to 40 K and beyond [3], leading to intensive efforts to further improve the $T$c.

It is known to be difficult to produce the ideal composition for superconductivity in metal-doped FeSe by the usual heating (annealing) method. Recently, a low-temperature route using liquid solvent has been found to achieve the intercalation of cations between the FeSe layers [4]. This method also provides an effective synthetic path for carbon-based materials [5,6] and layered inorganics [7]. Several solvents, such as $NH_3$, $(CH_3)NH_2$, and pyridine, can perform the metal dissolution, and improved superconductivity can be developed in metal-doped FeSe using this method [4,8]. Moreover, the solution method has achieved the co-intercalation of metal atoms and molecules in $Li_x(NH_3)_yFe_2Se_2$ [9]. Understanding the role that intercalated molecules play in structuring the electronic state may be the key to clarifying the mechanism of superconductivity in these materials. Among the metal-intercalated FeSe compounds synthesized by the low-temperature solution method, the structure of the superconducting phase in Ba-doped FeSe is still a matter of debate [4]. No report has been published on the synthesis of this compound by the usual high-temperature heating method. Ying et al. proposed no intercalation of $NH_3$ molecules in Ba-doped FeSe in the structure of the superconducting phase [4]. In this Letter, we describe the composition of a Ba-FeSe phase using the solution method and the determination of the optimum conditions to produce a high-Tc superconducting phase. We emphasize that this low-temperature route has the potential to control the production of excess iron. We also propose a crystal structure for the superconducting phase of Ba-doped FeSe in which both Ba and $NH_3$ molecules are co-intercalated between the FeSe layers.

We synthesized Ba-doped FeSe using liquid ammonia as the solvent. The high purity of the tetragonal β-FeSe phase is important because the shielding fraction of the Ba-doped sample was found to depend on the fraction of β-FeSe. First, we synthesized FeSe powder samples with a high superconducting fraction in accordance with previous reports [10], then placed the β-phase enriched FeSe powder sample and Ba metal in a glass tube. We treated the samples in an Ar-filled glove box with no exposure to air. The glass tube containing the powdered Ba and FeSe was connected to a vacuum line and dynamically pumped down to $10^{-2}$ Pa. Then $NH_3$ gas was condensed in the glass tube by cooling with liquid $N_2$. The glass tube was filled with liquid $NH_3$ ( ~ 10 ml) and was sealed. The solution was mixed with a stirrer for several hours at room temperature. The color of the solution changed to blue when the Ba metal dissolved (see figure 1(a)), changing to colorless and transparent when the reaction was complete. Next, we removed the liquid $NH_3$ by heating and dynamically pumping the glass tube. Finally, we obtained the Ba-doped powder samples. X-ray diffraction patterns were measured with a RIGAKU TTR-III to investigate the doping dependence of lattice parameters and the sample stoichiometry. To determine the structure of the superconducting phase, we used synchrotron radiation (KEK-PF, BL-8A) to obtain the X-ray diffraction pattern of the powder, with a wavelength of 1 Å selected to reduce absorbance by the sample. A Rietveld analysis was performed with the GSAS program package to determine the structure of the superconducting phase. The magnetization, $M$, was measured with a SQUID magnetometer (MPMS2, Quantum Design) in the temperature region > 2 K.

The quantitative composition of the samples was checked by the SEM/EDX analysis (VE-9800SP, Keyence Co. Ltd.). Figure 1 (b) shows the ratio of Ba to Se (Ba/Se) and Fe to Se (Fe/Se) as a function of $x$. Below $x = 0.8$, the Ba/Se ratio remained almost unchanged even when the nominal Ba concentration changed. However, above $x = 0.9$, the Ba/Se ratio suddenly increased. Figure 1(c) shows the XRD patterns with different $x$ -values. For comparison, we plotted the XRD pattern of the β-FeSe phase. Below $x = 0.4$, we can see traces of the diffraction peaks of the β-FeSe phase (e.g. around 2θ ~ 15 degree), as marked by the red arrow. As the Ba concentration increases, the β-FeSe phase disappears and new diffraction peaks (2 θ ~ 12 degree) are observed (blue arrow). As shown later, these new peaks correspond to the emergence of the superconducting phase. We note that the lattice parameters show a weak Ba concentration dependence [11]. Above $x = 0.9$, additional new peaks were found around 2 θ ~ 18 degrees (see the dotted circle in the figure). From the SEM/EDX and XRD experiments, as the Ba/Se concentration in the crystal is almost constant between $x = 0.3$ and 0.8, we conclude that stoichiometric samples are obtained in this region.

Figure 2(a) shows the temperature dependence of the magnetic susceptibility for different Ba concentrations. We can observe clear drops in the magnetic susceptibility around 38 K for Ba-doped samples. The inset shows the magnetic susceptibility for the zero-field cooling condition (ZFC) and for the field cooling condition (FC). This result implies a superconducting transition for the Ba-doped samples. The observed superconducting transition temperature ($T$c) corresponds closely to the previous report [4]. Figure 2 (b) summarizes the $T$c and the shielding fraction as a function of the nominal value of $x$. Below $x = 0.3$, a small amount of the superconducting phase of β-FeSe coexisted, but the superconducting fraction suddenly decreased above $x = 0.4$ and finally disappeared. The $T$c for Ba-doped samples depended weakly on the $x$-value, and the highest $T$c was obtained around $x = 0.5$. However, the shielding fraction showed a different $x$-value dependence. The shielding fraction decreased above $x = 0.8$ while the $T$c remained unchanged. As can be seen in figure 2(a), the normal state susceptibility is enhanced with increasing $x$. Above $x = 0.8$, the Curie-Weiss law was followed, and the Curie constant increased with increasing $x$. XRD showed additional new peaks above $x = 0.9$. SEM/EDX analysis also showed a change in the Ba/Se ratio in the same region of $x$. Therefore, an impurity phase is produced above $x = 0.9$, and its paramagnetism should suppress the shielding fraction of the superconducting phase. From the $x$-dependence of the superconducting properties, we conclude that the optimum value of $x$ is limited to the region from 0.3 to 0.7. The highest $T$c (38.8 K) is obtained at $x = 0.5$ and the maximum shielding fraction currently observed is greater than 48 %. Moreover, we reproduced the observation that the shielding fraction depends on the time interval following sample preparation, which was also noted by Ying et al. [4]. As shown later, we include the intercalation of $NH_3$ molecules in the structural model of the superconducting phase. Thus, the time evolution of the shielding fraction may be due to the removal of $NH_3$ molecules from the crystal.

To identify the structure of the Ba-doped FeSe superconducting phase, we examined two structural models for the superconducting phase, one with $NH_3$ molecules included between the FeSe layers ($Ba_x(NH_3)_yFe_2Se_2$) and the other without ($Ba_xFe_2Se_2$). For the former case, we placed the $NH_3$ molecules at the same position as in the $Li_x(NH_3)_yFe_2Se_2$ structure [9]. We used the $x = 0.4$ sample as shown in Fig. 1(c) [11] and compared the structural refinement factor ($R$-factor) for two models. Ying et al. discussed the possibility of $NH_3$ intercalation between the FeSe layers in the structure of the superconducting phase [4], and claimed that the intercalation of $NH_3$ molecules gave no improvement in the $R$-factor. However, our structural refinements using the synchrotron radiation data clearly indicate an improvement in the $R$-factor due to the intercalation of $NH_3$ molecules between FeSe layers. Figure 3 shows the XRD data and the Rietveld refinements with three structural phases (doped phase, tetragonal β-FeSe, and hexagonal FeSe). The pink, light blue and dark blue vertical bars indicate the candidate diffraction peaks for these three phases, respectively, and the bottom line indicates the difference between the calculated values and the raw data. During the refinements, the fractions of the three phases were found to be 61% for $Ba_x(NH_3)_yFe_2Se_2$, 29% for tetragonal β-FeSe and 10% for hexagonal FeSe, respectively. The $R_w$-factor and $R$-factor for the $Ba_xFe_2Se_2$ case and the $Ba_x(NH_3)_yFe_2Se_2$ case were 4.4% and 3.3%, and 3.6% and 2.6%, respectively. We also changed the occupancy of Ba and Fe atoms in this refinement. As the final model refinement with $NH_3$ molecules yields an improved $R$-factor, we conclude that the structure of the superconducting phase of Ba-doped FeSe includes $NH_3$ molecules in a way similar to $Li_x(NH_3)_yFe_2Se_2$. Figure 3(b) shows the structural model of the superconducting phase of Ba-doped FeSe, and the refined structural parameters are described in Fig.3(a). From the occupancy of Ba and Fe atoms, we estimate that the composition of the sample is $Ba_{0.4}(NH_3)_{1.0}Fe_{1.8}Se_{2.0}$, which is consistent with the SEM/EDX elemental analysis.

The present experimental results support the co-intercalation of Ba and $NH_3$ molecules between the FeSe layers. The quantities of Ba and $NH_3$ molecules intercalated between the FeSe layers are both weakly dependent on $x$, possibly leading to the weak dependence on $x$ of the $T_c$ values. Thus, controlling the quantity of intercalated metal atoms and molecules is a key to increasing $T_c$ in the present Ba-FeSe system. It is also known that Ba-doped FeSe is difficult to synthesize by the usual heating (annealing) method. One of the roles of $NH_3$ molecules is to enlarge the space between the FeSe layers. The factors controlling the formation of crystal structure remain open to argument. Additional theoretical information will be needed to clarify the role of intercalated molecules in determining the electronic state.

In the structure of the superconducting phase, the Fe-Fe distance ($d_{Fe-Fe}$) between neighboring FeSe layers is estimated to be 7.9709(18) Å. It is suggested that the $T_c$ in the metal-doped FeSe system increases with increasing distance between neighboring FeSe layers [10,12]. Our experimental evidence ($T_c$ = 38.8 K and $d_{Fe-Fe}$) is quantitatively consistent with this. Accordingly, the expansion of the distance between the FeSe layers by the intercalation of larger molecules may provide a new way to increase $T_c$ in metal-doped iron selenides.

We would like to thank Prof. M. Nohara and Dr. K. Kudo for stimulating discussions on superconductivity and for the use of their glove box to synthesize the FeSe powder sample. This research is partly supported by Grant-in-Aid No. 23340104 from MEXT, Japan.

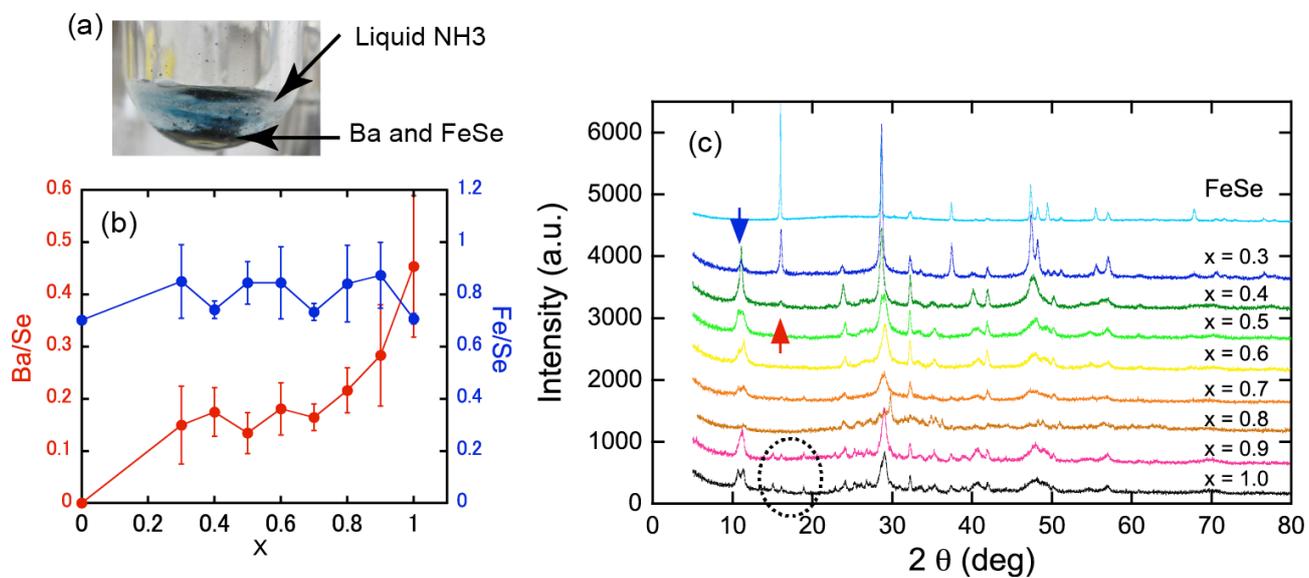

Figure 1. (a) Photo of liquid NH$_3$ solution with Ba and FeSe powder while reacting (b) Ba/Se and Fe/Se ratios as a function of *x*, estimated by the SEM/EDX method (c) XRD patterns with different Ba concentrations

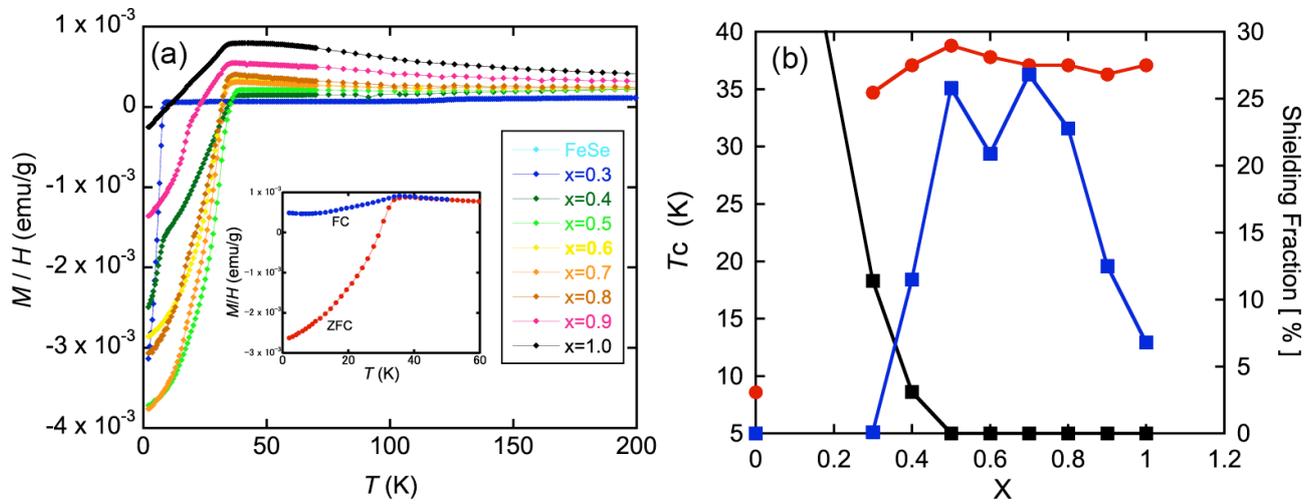

Figure 2. (a) Temperature dependence of magnetization as a function of *x*. The inset shows the temperature dependence of magnetization under field-cooled (FC) and zero-field cooled (ZFC) conditions. (b) Superconducting transition temperature $T_c$ (red circle) and shielding fraction for β-FeSe (black square) and the Ba-doped phase (blue square) as a function of *x*.

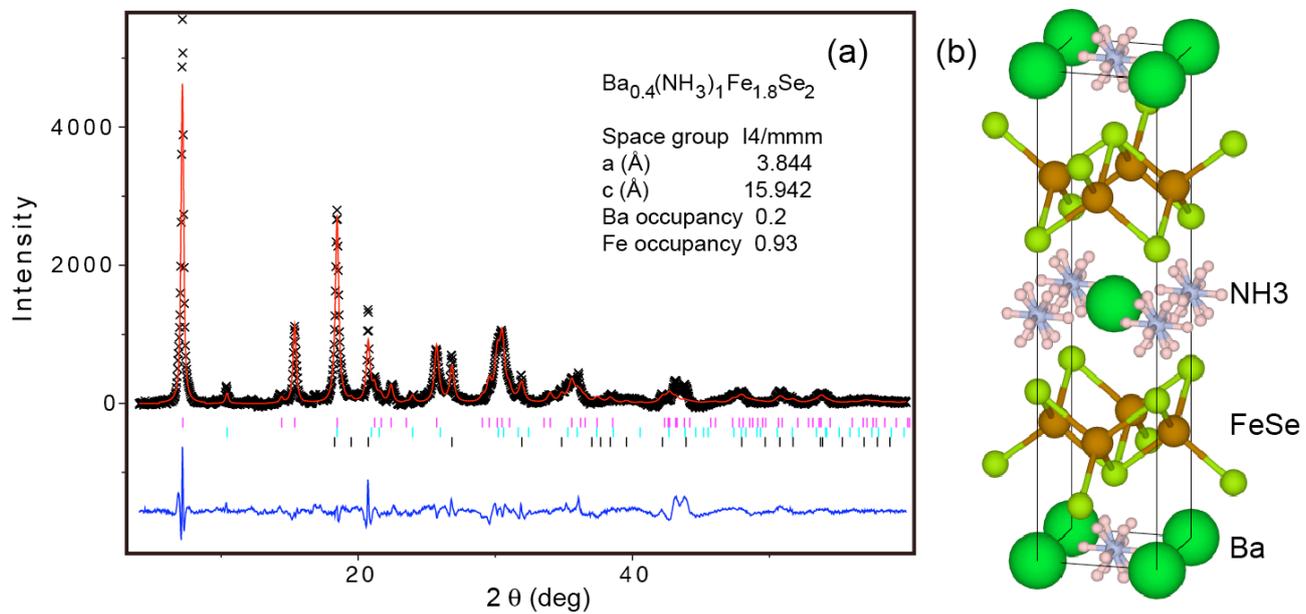

Figure 3. (a) XRD pattern with synchrotron radiation ($\lambda =1$ Å) and refined lattice parameters. The Rietveld refinement was performed using three structural phases (tetragonal β-FeSe, hexagonal FeSe, and doped phase). (b) Structural model for the superconducting phase of Ba-doped FeSe.

References

1. F. C. Hsu, J. Y. Luo, K. W. The, T. K. Chen, T. W. Huang, P. M. Wu, Y. C. Lee, Y. L. Huang, Y. Y. Chu, D. C. Yan, M. K. Wu, Proc. Natl. Acad. Sci. U.S.A. 105, 14263 (2008).

2. A. Subedi, L. Zhang, D. J. Sing, M. H. Du, Phys. Rev. B 78, 134514 (2008).

3. J. G. Guo, S. Jin, G. Wang, S. Wang, K. Zhu, T. Zhou, M. He, X. Chen, Phys. Rev. B 79, 180520(R) (2010).

4. T. P. Ying, X. L. Chen, G. Wang, S. F. Jin, T. T. Zhou, X. F. Lai, H. Zhang, W. Y. Wang, Sci. Rep. 2, 426 (2012).

5. A. Y. Ganin, Y. Takabayashi, Y. Z. Khimyak, S. Margadonna, A. Tamai, M. J. Rosseinsky, K. Prassides, Nat. Materials, 7, 367 (2008).

6. T. Kambe, X. He, Y. Takahashi, Y. Yamanari, K. Teranishi, H. Mitamura, S. Shibasaki, K. Tomita, R. Eguchi, H. Goto, Y. Takabayashi, T. Kato, A. Fujiwara, T. Kariyado, H. Aoki, Y. Kubozono, Phys. Rev. B 86, 214507 (2012).

7. R. B. Somoano, V. Hadek, A. Rembaum, J. Chem. Phys., 58, 697 (1973).

8. A. Krzton-Maziopa, E. V. Pomjakushina, V. Yu. Pomjakushin, F. von Rohr, A. Schilling, K. Conder, J. Phys.:Cond. Matter, 24, 382202 (2012).

9. M. Burrard-Lucas, D. G. Free, S. J. Sedlmaier, J. D. Wright, S. J. Cassidy, Y. Hara, A. J. Corkett, T. Lancaster, P. J. Baker, S. J. Blundell, S. J. Clarke, Nat. Materials, 12, 15 (2013).

10. A. Zhang, T. Xia, K. Liu, W. Tong, Z. Yang, Q. Zhang, Sci. Rep. 3, 1216 (2013).

11. As shown in figure 1(c), the peaks around (002) index ($2\theta \sim 12$ degree) may have little splitting, suggesting a distribution of the length of $c$-axis in the doped samples. In the structural refinements of the superconducting phase, we selected the sample with no splitting of this peak.

12. D. Liu, W. Zhang, D. Mou, J. He, Y. Ou, Q. Wang, Z. Li, L. Wang, L. Zhao, S. He, Y. Peng, X. Liu, C. Chen, L. Yu, G. Liu, X. Dong, J. Zhang, C. Chen, Z. Xu, J. Hu, X. Chen, X. Ma, Q. Xue, X. J. Zhou, Nat. Commun. 3, 931 (2012).